\documentclass[twocolumn]{aastex631}
\usepackage{amsmath}
\shorttitle{Origin of line broadening in fading granules}
\shortauthors{Ishikawa et al.}
\graphicspath{{./}{figures/}}

\begin{document}

\title{Origin of line broadening in fading granules:
influence of small-scale turbulence}

\author[0000-0002-4669-5376]{Ryohtaroh T. Ishikawa}
\affiliation{National Institute for Fusion Science \\
322-6 Oroshi-chi, Toki, Gifu 509-5292, Japan}
\affiliation{Fusion Science Program, Graduate Institetu for Advanced Studies, The Graduate University for Advanced Studies, SOKENDAI\\
322-6 Oroshi-chi, Toki, Gifu 509-5292, Japan}

\author[0000-0002-5054-8782]{Yukio Katsukawa}
\affiliation{National Astronomical Observatory of Japan \\
2-21-1 Osawa, Mitaka, Tokyo 181-8588, Japan}
\affiliation{Astronomical Science Program, Graduate Institetu for Advanced Studies, The Graduate University for Advanced Studies, SOKENDAI\\
2-21-1 Osawa, Mitaka, Tokyo 181-8588, Japan}



\begin{abstract}
In the quiet region of the solar photosphere, turbulent convective motions of the granular flows naturally drive the subgranular-scale flows. However, evaluating such small-scale velocities is challenging because of the limited instrumental resolution. Our previous study, \citet{RTI20a}, found line broadening events during fading process of granules; however, their physical mechanism has remained unclear. In the present study, we observed the fading granules with the Hinode-SOT/SP and performed spectral line inversions. Moreover, we investigated broadening events of synthesized spectra in fading granules reproduced by the MURaM simulation. Our results demonstrated that the small-scale turbulent motions are excited in the fading process and such turbulent flows contribute to line broadening. The spectral line widths can be potential tracers of the photospheric turbulent flows.

\end{abstract}



\section{Introduction} \label{sec:intro}
The quiet region of the solar photosphere is covered by numerous cellular patterns, termed granules. The granular motions are so turbulent that the granulation exhibits various dynamics such as birth, death, merger, and fragmentation \citep{Hirzberger99a,Lemmerer17}. The turbulent nature of granulation is also represented by the broad feature in the power spectra of kinetic energy \citep{Rieutord10,Katsukawa12} as well as by the power-law component of the probability density function of size of granules \citep{Abramenko12}. 
The turbulent nature inherently drives the subgranular-scale flows via a cascading process resulting from the energy injection at the granular scale.
Such subgranular-scale flows are critical in creating small-scale magnetic structures through a local dynamo mechanism, and more than half of the total magnetic energy is assumed to be hidden on spatial scales smaller than 100 km \citep{Rempel14}. This 
spatial scale has remained difficult to resolve.

The microturbulence velocity is one of the ways to describe the unresolved velocity fields.
The microturbulence term was originally introduced to explain the large equivalent widths that were unexplainable only with thermal broadening.
Although some inversions use the microturbulence term, the physical mechanism that produces such excess broadening is still unclear.

The extent of the microturbulence term necessary for explaining the photospheric line profiles observed with a high spatial resolution,
which are associated with uncertainties in small-scale velocities, remains unclear. \citet{Socasnavarro11} assumed that the small-scale velocity corresponding to the microturbulence term is so small that the spectral line profiles observed with Hinode-SOT can be explained without microturbulence. \citet{BellotRubio05} and \citet{QuinteroNoda14b} also explain the photospheric spectra obtained with Hinode-SOT without the microturbulence term, which demonstrates that the small-scale velocity field is much smaller than the thermal velocity $\sqrt{2k_BT/m}=1.3$ km/s at $T=6000$ K. Some other studies, however, observed that the microturbulence term was necessary in the photosphere \citep{LopezAriste07,Buehler15,Guglielmino20}. \citet{Buehler15} estimated the average microturbulent velocities of 3.1 km/s at $\log\tau=0$ and 0.8 km/s at $\log\tau=-0.9$ by performing a spectral line inversion with a deconvolution technique that reduced the effect of the spatial resolution of Hinode-SOT. These values were comparable to the root-mean-square velocity related to granulation of approximately 1.1 km/s \citep{Oba17b}. This implies that the velocity fields on small scales are comparable to those on granular scales.

Small-scale variations in LOS velocity are measured by analyzing the spectral line broadening as mentioned above. The spatial distributions of the FWHM of photospheric spectra are observed to be larger in intergranular lanes than in the granules using observations barely resolving the granulation scale \citep{Holweger89,Nesis92}.
\citet{Nesis92} stated that line broadening is related to turbulent motions caused by shocks driven by supersonic flows in granulation. However, \citet{Solanki96} and \citet{Gadun97} performed 2-dimensional hydrodynamic numerical simulations
and calculated the photospheric spectral line widths. They concluded that LOS gradients of LOS velocities related to granulation can cause spectral line broadening and excess line broadening in intergranular lanes can be explained
using the strong gradient of vertical velocities. In addition, \citet{Solanki96} and recent numerical simulation by \citet{Vitas11} demonstrated that the supersonic motions are preferentially seen as horizontal flows, which implied that line broadening relevant to shocks can be observed only in a limb observation \citep{BellotRubio09}.

\citet{RTI20a} found that the excessive spectral line broadening appeared in spatially localized regions with the Hinode-SOT/SP.
Some of them were associated with the fading process of granules.
Velocity difference between the upper and lower photosphere was enhanced in the fading phase of granules and simultaneous line broadening was also observed.
They concluded that these line broadenings cannot be explained only by the Doppler velocity gradient obtained with the bisector analysis due to the small-scale fluctuations of LOS velocity
 in fading granules.

There is a limitation to estimating the Doppler velocity gradient using the bisector analysis. The Hinode-SOT/SP observed two neutral iron lines 6301.5 {\AA} and 6302.5 {\AA}. The bisector analyses only analyzed the 6301.5 {\AA} line and used the two bisector velocities to calculate the velocity difference between the lower and upper photosphere, providing an approximate estimation of velocity gradient along the LOS. To examine the origin of the line broadening quantitively,
the contributions of temperature and velocity by the LOS gradients to the line broadening should be evaluated by analyzing the entire profiles of the spectra based on the radiative transfer. Therefore, spectral line inversions were performed to determine the detailed atmospheric structures in the granulation. Moreover, we analyzed MHD simulation data and synthesized spectra to reveal the photospheric dynamics that caused the spectral line broadenings.

\section{Observation and Method}
\subsection{Observation}
We observed the quiet region near the disk center with a spectropolarimeter (SP; \citealt{Lites+13}) of the Solar Optical Telescope \citep{Tsuneta08} aboard the Hinode spacecraft \citep{Kosugi07}.
SOT achieves a spatial resolution of 0.3 arcsec (corresponding to 200 km on the solar surface) owing to the diffraction-limited performance at 6300 {\AA}.

We analyzed the dynamic mode data obtained on November 8, 2018.
This dataset was used in \citet{RTI20a}.
In this observation, SOT repeats 15-position raster scans, covering $2^{\prime\prime}.3\times 82^{\prime\prime}$ field-of-view (FOV), with a time cadence of 17.8 s.
The plate scale of a pixel is $0^{\prime\prime}.15\times 0^{\prime\prime}.16 = 0.11 \mathrm{Mm} \times 0.12 \mathrm{Mm}.$
The slit is directed in the North-South direction and the scan direction in East-West.
The average root-mean-square fluctuation of the continuum intensity $\sigma_I$ of 1.0\% is an expected photometric error from the data,
which was evaluated in the wavelength range from 6301.83 {\AA} to 6302.17 {\AA},
where the continuum spectra between the two iron lines was observed.

\begin{table}[t]
  \begin{center}
    \caption{Number of nodes for the inversions}
    \begin{tabular}{c|c|c}
                 & Without microturbulence & With microturbulence\\ \hline
     Temperature &     5 & 5\\
     Doppler Vel.&    5 & 5\\
     microturbulence&  0  &  1\\
     $N_n$& 10 & 11
    \end{tabular}
    \label{tb:cond_inv}
  \end{center}
\end{table}

\subsection{Observational Parameters}
We obtained some observational parameters to quantitatively capture the features of the observed spectral line shapes (see Table 2 of \citealt{RTI20a}). We focused on four parameters such as the continuum intensity ($I_{\mathrm{cont}}$), the FWHM, the velocity difference ($\Delta v$), and the equivalent width.
The velocity difference are derived by th bisector analysis and is defined by $\Delta v = v_{0.05}-v_{0.7}$,
where $v_{0.05}$ and $v_{0.7}$ are the bisector velocities at the normalized intensities of 0.05 and 0.7.
We can evaluate the asymmetric shape of the observed spectral line profile with $\Delta v$ which should be caused by the Doppler velocity variation along the LOS.

\subsection{Spectral Line Inversion}\label{sec:method_inversion}
We perform spectral line inversion using the SIR code
(Stokes Inversion based on Response function; \citealt{Ruizcobo92}).
SIR code can fit the observed spectra with a 1-dimensional atmosphere.
Response function $\bm{R}\bm{R}_{\xi}$ is based on a first-order perturbative analysis of the radiative transfer equation
for polarized light \citep{Landi85,SanchezAlmeida92}
and is defined as
\begin{equation}
\delta \bm{I}(\lambda) = \int_0^{\infty} \bm{R}_{\xi}(\lambda, \tau)\delta \xi(\tau) d\tau, \label{eq:response_function}
\end{equation}
where $\delta \xi(\tau)$, $\tau$, and $\bm{I}$ are perturbation of a physical quantity,
optical depth at 5000 {\AA}, and Stokes profiles normalized with the average continuum intensity,
respectively \citep{Landi77}.
The response function represents the linear relationship
between the perturbation and
corresponding small change in the emergent Stokes spectra $\bm{I}(\lambda)$.
The continuum intensity reflects the temperature at approximately $\log\tau=0$,
and both Fe I 6301.5 {\AA} and 6302.5 {\AA} lines have a sensitivity to the temperature from $\log\tau=0$ to $\log\tau=-3$.
The lines are also sensitive to the Doppler velocity and microturbulence at approximately $\log\tau=-1$.

To investigate the small-scale turbulent motions,
we inverted the observed line profiles both with and without microturbulence and compared the results.
The number of nodes of the inversions is listed in Table \ref{tb:cond_inv}.
The optical depths at which perturbations are sought are determined only by the number of nodes,
which are logarithmically equally spaced.
In this study, we considered the atmosphere from $\log\tau=1.0$ to $\log\tau=-3.8$,
putting the nodes at $\log\tau=1.0, -0.2, -1.4, -2.6, -3.8$
for the temperature and Doppler velocity.
After the perturbations were introduced to the parameters at each node,
they were interpolated with cubic-spline functions.
In the inversion without microturbulence, the microturbulence term is fixed with 0 km/s,
while it is a free parameter but assumed to be constant along the LOS in the inversion with microturbulence.
The line-spread-function of the Hinode-SOT/SP \citep{Lites+13} is considered.

The inversion was done only for the Stokes {\it I} spectra and the magnetic field strength was assumed to be zero,
because the concerned region has small total polarization $P_{\mathrm{tot}}<1.6$\% that was evaluated using the \verb|sp_prep| routine \citep{Lites_Ichimoto13}.
Goodness of the fitting  was evaluated with a reduced $\chi^2$ defined as
\begin{equation}
  \chi^2 \equiv \frac{1}{N_{\lambda}-N_n} \sum_{\lambda_i=1}^{N_{\lambda}}
  \left| \frac{I_{\mathrm{obs}}(\lambda_i)-I_{\mathrm{fit}}(\lambda_i)}{I_{\mathrm{c}}\sigma_I} \right|^2. \label{eq:chi2}
\end{equation}
Here $N_{\lambda}$, $N_n$, and $\sigma_{I}$ indicate the number of pixels for the wavelength,
number of free parameters, and photometric error
of the observed spectra, respectively.

The SIR code performs the Levenberg-Marquardt algorithm to determine the solution
that minimizes the $\chi^2$
by calculating the derivative of $\chi^2$ with the response function.
However, it still has a dependence of the inversion results on the initial parameters
and the solution is trapped in a local mininum.
To avoid such dependence,
we inverted each spectrum 2500 times with random initial guesses
and adopted the inversion result, which achieved the minimum $\chi^2$.
The initial values of temperature ranged between $\pm1500$ K from the Harvard-Smithsonian reference atmosphere (HSRA;\citealt{Gingerich71}) at each node,
that of Doppler velocity ranged from -2 km/s to 2 km/s at each node,
and those of microturbulence from 0 km/s to 2 km/s.

\edit2{\section{Results}} \label{sec:results}
To investigate the line broadening mechanism related to the granulation,
we analyzed two fading granules observed in the dynamic mode data.
Observed spectra in both fading granules had wide spectral line profiles and small $P_{\mathrm{tot}}<1.6$\%.
Owing to the small polarization, the line broadening of these spectra
is unlikely to be a result of the Zeeman effect.

\begin{figure*}[p]
\begin{center}
\includegraphics[width=12cm]{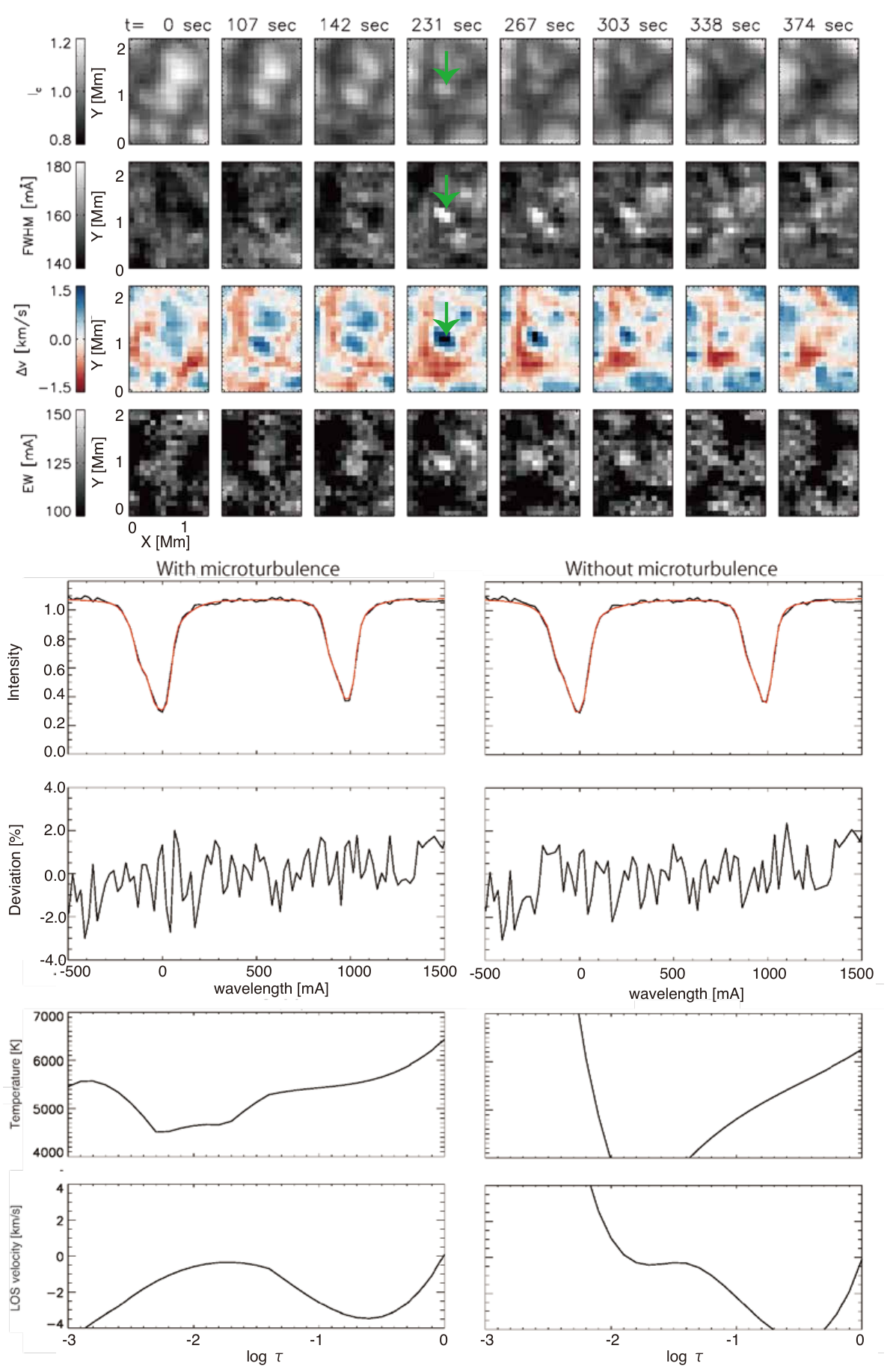}
\end{center}
\caption{The temporal evolution of a fading granule.
  Upper three rows show the temporal evolutions of continuum intensity, FWHM of Fe I 6301.5 {\AA} line, the velocity difference $\Delta v$, and the equivalent width of Fe I 6301.5 {\AA} line from top to bottom.
  The observed spectrum and inversion results at (x,y)=(0.6, 1.1) Mm at t=231 s are also shown.
  The observed Fe I 6301.5 {\AA} spectrum has FWHM of 192 m{\AA} and $\Delta v$ of 1.56 km/s.
  Bottom left panels show inversion results with microturbulence, inferring the microturbulence of 1.3 km/s.
  Bottom right panels show results without microturbulence.
  The reduced $\chi^2$ values are 1.34 with microturbulence and 1.42 without microturbulence.}
\label{fig:inv_fading_time}
\end{figure*}

Figure \ref{fig:inv_fading_time} shows the first case we analyzed (hereafter sample 1),
which was the same event analyzed by \citet{RTI20a}.
The top four rows show the temporal evolutions of the bisector parameters
during the fading process of the granule:
normalized continuum intensity ($I_{\mathrm{c}}$), spectral line widths (FWHM),
bisector velocity difference between upper and lower photosphere ($\Delta v$),
and the equivalent width (EW) of Fe I 6301.5 {\AA} line
defined in \citet{RTI20a}.
Line broadening was seen at around $(x,y)=(0.6, 1.1)$ Mm.
The Fe I 6301.5 {\AA} line profile at this pixel had FWHM of 192 m{\AA}, $\Delta v$ of 1.56 km/s and EW of 155 m{\AA}, which are largely different from the values appear in other pixels.
These temporal evolutions of the observational parameters show a sporadic change in the spectral line profile, which suggests small-scale dynamics in the photosphere.

Lower panels in Figure \ref{fig:inv_fading_time} showed the inversion results
for the center of the fading granule at $(x,y)=(0.6, 1.1)$ Mm at $t=231$ s.
Both the inversions with and without microturbulence successfully reproduced the observed spectral line profiles
with $\chi^2$ of 1.34 with microturbulence and 1.42 without microturbulence.
However, the estimated atmospheres were completely different.
The inversion with microturbulence obtained an atmosphere with the microturbulent velocity of 1.3 km/s,
while that without the microturbulence required large gradients of temperature and Doppler velocity
for reproducing the wide spectral line width.
The inverted profiles of temperature are also different.
In the absence of a microturbulent term the line broadening is fitted by
a strong rise of temperature in higher layers.

\begin{figure*}[p]
\begin{center}
\includegraphics[width=12cm]{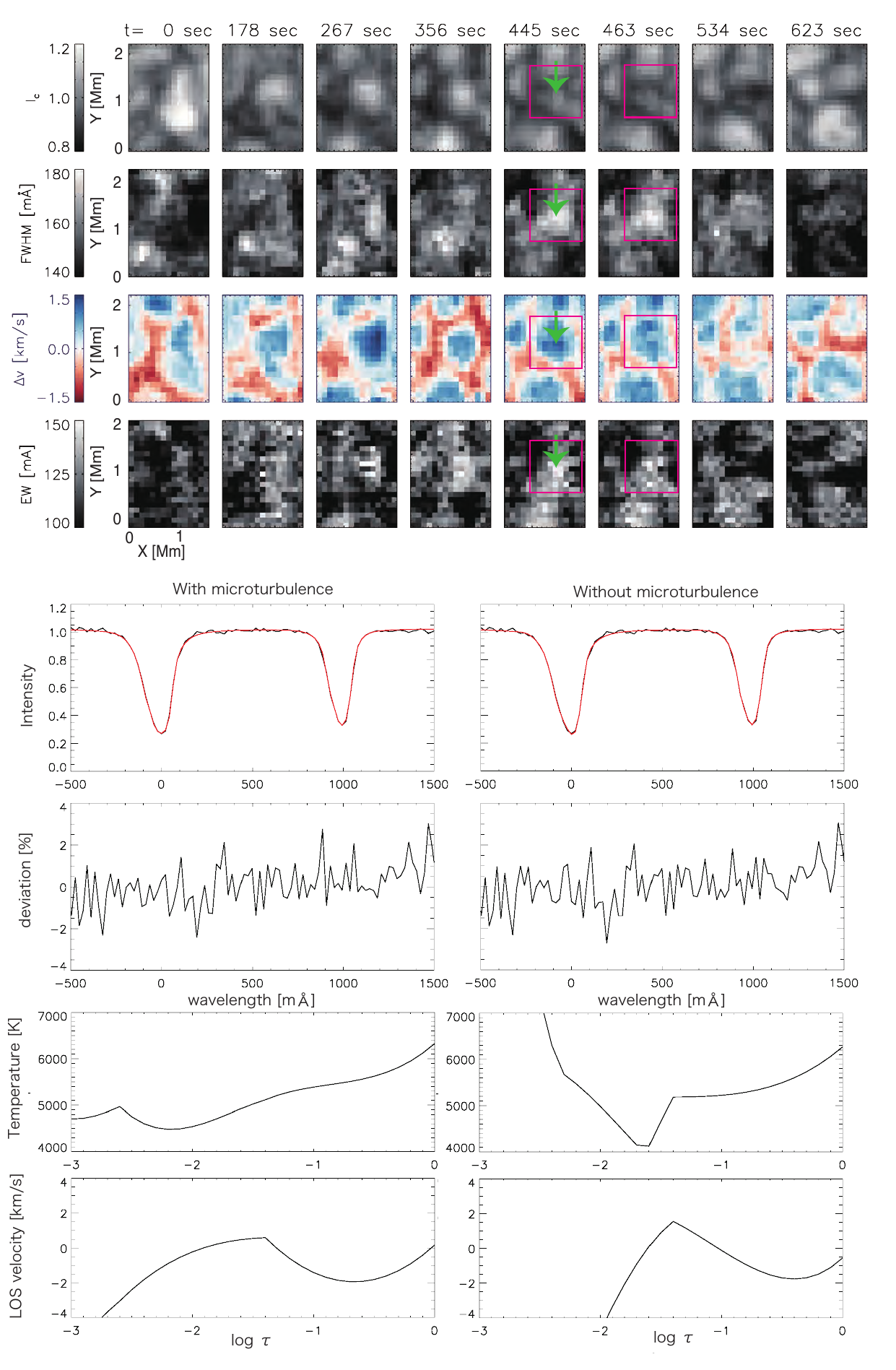}
\end{center}
\caption{Same figure as Figure \ref{fig:inv_fading_time} except for sample 2.
  The observed spectrum and the inversion results at (x,y)=(1.1, 1.2) Mm at t=445 s (indicated by green arrows) are also shown.
  The observed Fe I 6301.5 {\AA} spectrum has FWHM of 175 m{\AA} and $\Delta v$ of 1.15 km/s.
  Bottom left panels show inversion results with microturbulence, inferring the microturbulence of 0.9 km/s.
  Bottom right panels show results without microturbulence.
  The reduced $\chi^2$ values are 1.06 with microturbulence and 1.07 without microturbulence.
  Inversion results inside the magenta boxes are shown in Figure \ref{fig:inv_temp}.}
\label{fig:inv_fading_time2}
\end{figure*}

\begin{figure*}[t]
\begin{center}
\includegraphics[width=18cm]{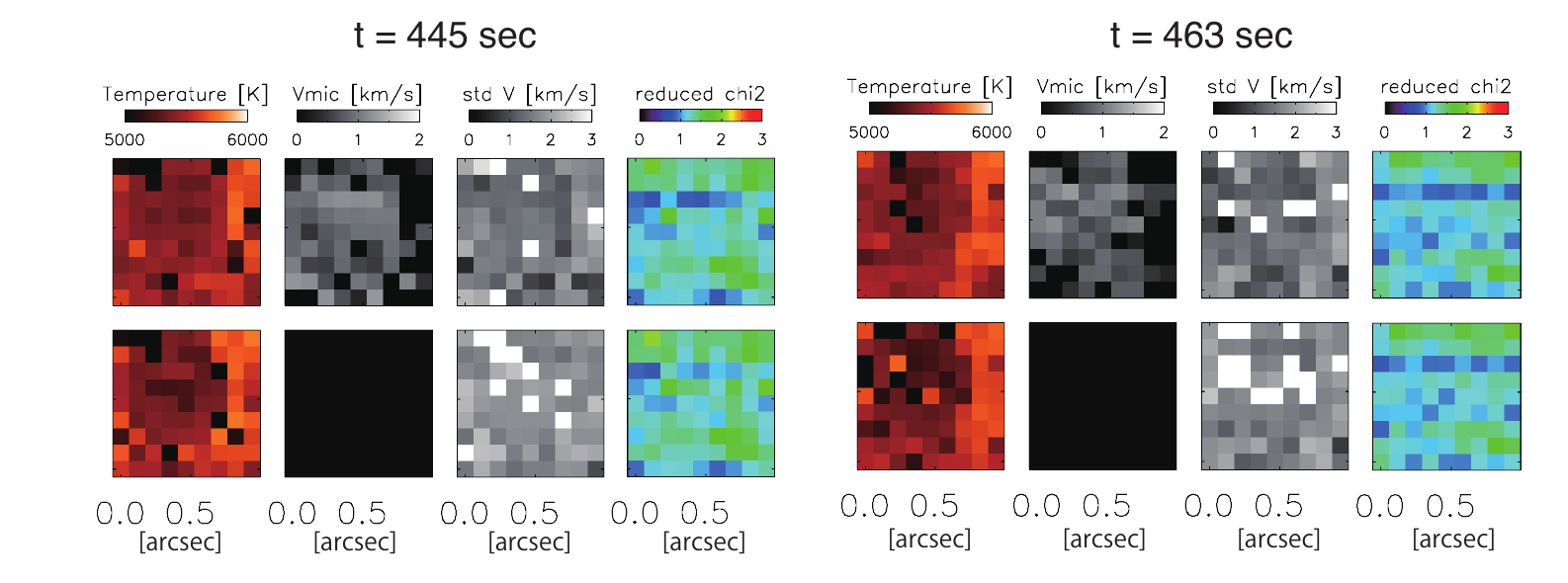}
\end{center}
\caption{The spatial distributions of inverted physical quantities of sample 2 (indicated by the magenta boxes in Figure \ref{fig:inv_fading_time2}) with and without microturbulence at $t=445$ s and 463 s when large FWHM is observed.
For the inversion results, temperature at $\log\tau=-1$, microturbulence velocity, standard deviation of LOS velocity in $-2.5\leq\log\tau\leq-0.5$, and the reduced $\chi^2$ are displayed.
The FOV of each image is 1 Mm $\times$ 1 Mm.}
\label{fig:inv_temp}
\end{figure*}

\begin{figure*}[tbp]
\begin{center}
\includegraphics[width=14cm]{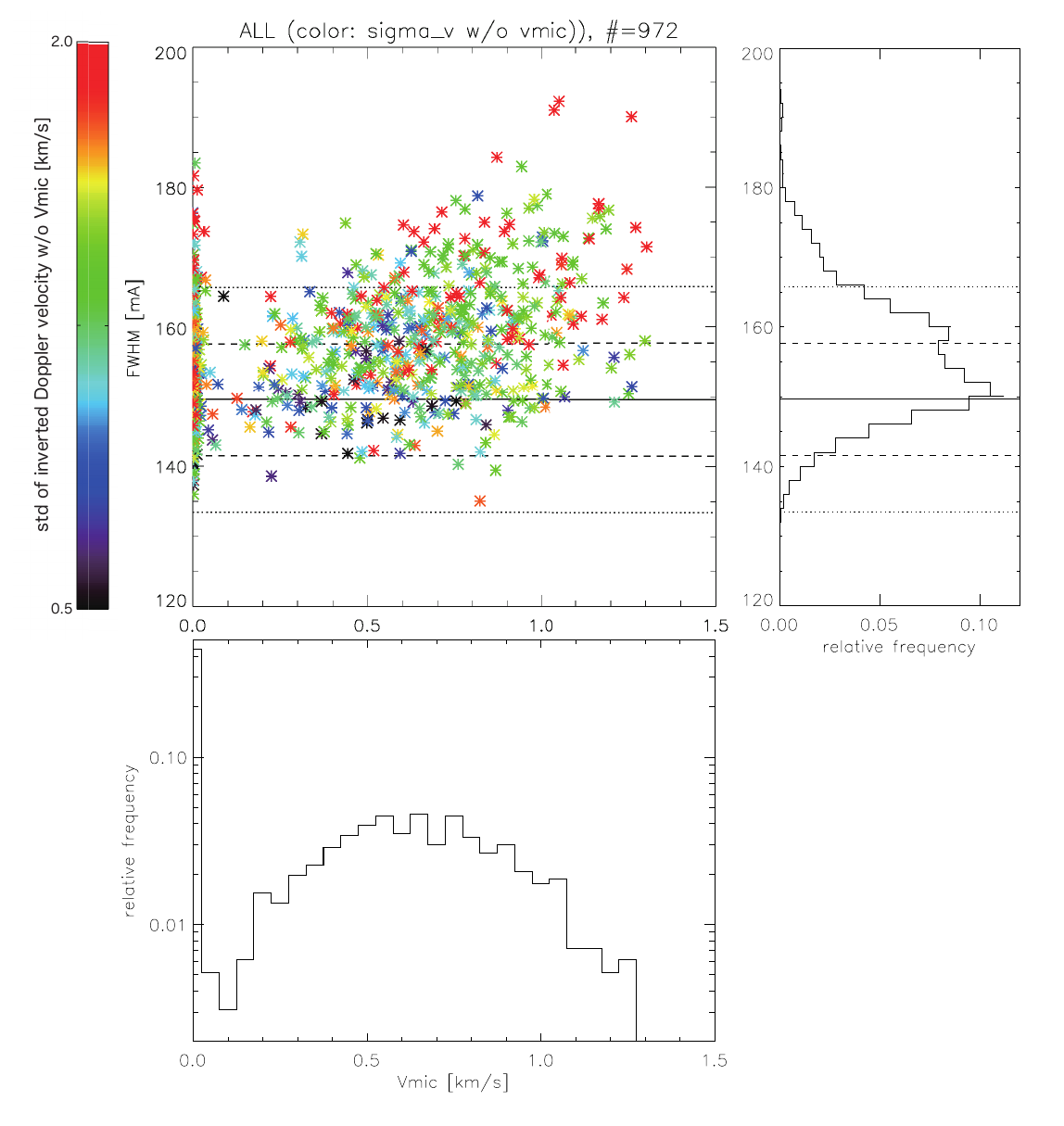}
\end{center}
\caption{Scatter plot of FWHM and the inverted microturbulent velocity
  of the local regions around fading granules.
  The color shows the standard deviation of the Doppler velocity at each pixel inferred
  by the inversion without microturbulence.
  The horizontal solid line shows the average FWHM measured in the data.
  The horizontal dashed and dotted lines indicate the FWHM $\pm 1\sigma$ and $\pm 2\sigma$, respectively.
  The histogram of the FWHM in the two regions inverted is also shown in the right panel,
  emphasizing that most of the inverted spectra have large FWHM
  as broad spectral line profiles are observed around fading granules.
  Bottom panel shows the histogram of inferred microturbulence velocity.}
\label{fig:fwhm_vmic}
\end{figure*}

Another line broadening event (sample 2) associated with a fading granule
is shown in Figure \ref{fig:inv_fading_time2} detected in the dynamic mode data.
Also in this event, both the inversion scenarios can reproduce the observed spectra,
estimating largely different atmospheric structures.
The inversion with microturbulence obtains the atmosphere with microturbulent velocity of about 0.9 km/s,
while the result without microturbulence requires large gradients of temperature and Doppler velocity
to reproduce the wide spectral line width.

Figure \ref{fig:inv_temp} shows the spatial distributions of inversion results
around a fading granule of sample 2.
The standard deviation of Doppler velocity was calculated from $\log\tau=-0.5$ to $-2.5$.
This standard deviation represents the velocity gradient on large scales because the small-scale velocity variation
cannot be detected
due to the interval of the locations of nodes of $\Delta(\log\tau)=1.2$
in the inversion.

The spatial distribution of inversion results were not smooth.
This was because the broad spectra were observed in a local region,
which was indicative of small-scale variations of physical quantities.
In addition, the $\chi^2$ values fitted with and without microturbulence were similar to each other, and 
even with microturbulence, the inversion sometimes adopts the large-gradient scenario
to explain the line broadening with a small microturbulence.

Figure \ref{fig:fwhm_vmic} shows the scatter plot of the FWHM and inferred microturbulent velocities
in the regions near fading granules of sample 1 and 2.
Color denotes the standard deviation of Doppler velocity at each pixel inferred
by the inversion without microturbulence.
We emphasized that these standard deviations indicated the velocity gradient on scales larger than
$\Delta(\log\tau)\gtrsim1.2$ that is the interval of the locations of the nodes.
On the other hand, the microturbulence term stands for velocity variations
on scales smaller than $\Delta(\log\tau)\lesssim1.2$.
Although there is a large scatter,
the microturbulence velocities of approximately 1 km/s are estimated from the wide spectra
with FWHM larger than 165.8 m{\AA}.
The large variation in Doppler velocity along the LOS is required
to reproduce such wide spectral line profiles without mictoturbulence.
The average standard deviation of Doppler velocity
with FWHM from 141.5 m{\AA} to 157.7 m{\AA} is 1.3 km/s,
while that with FWHM larger than 165.8 m{\AA} is 3.2 km/s.
Notably, the average FWHM obtained with the entire FOV of the normal map data was 150 m{\AA}.
The inversion was performed in local regions in the vicinity of fading granules,
preferentially showing larger FWHM than average.
These results demonstrate that
the variation in  Doppler velocity along the LOS can explain the spectral line broadening
without the microturbulence.
In the inversion with microturbulence,
some results have small microturbulence velocity for large FWHM.
These results also explain the large FWHM by including the variation in Doppler velocity along the LOS.
Therefore, the two scenarios cannot be distinguished 
only with the two neutral iron lines observed with Hinode-SOT/SP.

The inverted microturbulence velocities have a bimodal distribution:
first is the solution with microturbulence larger than 0.2 km/s
and other is the solution with microturbulence near zero
as shown in the bottom panel in Figure \ref{fig:fwhm_vmic}.
Only a few pixels have microturbulence of approximately 0.1 km/s.

\section{Discussion of the Inversion Results}
\label{sec:discussion_inv}
Our results show the difficulty in estimating the height variation of Doppler velocities from wide spectral line profiles of the two neutral iron lines observed by Hinode-SOT/SP.
One of the causes of the difficulty is the limited range of the formation height of the lines.
Both the lines are sensitive to Doppler velocities in the middle photosphere (at around $\log\tau=-1$)
and had low sensitivity to those in other layers.
Multi-line observation and simultaneous inversion can be a powerful way to resolve this problem.
For example, the neutral iron line at 1.56 {\textmu}m that is formed in the bottom photosphere may have a strong capability to constrain the velocity profiles.

Another issue is possibly associated with the definition of the microturbulence term.
The LOS velocity and the microturbulence term are formulated as separate terms.
In terms of the hydrodynamics, however, their difference is just at a spatial scale and they are seamlessly connected.
If there are velocity variations on scales smaller than the finite width of the response function,
the spectral line profile and the bisector analysis cannot resolve them and the velocity variations may play a similar role as the microturbulence term.
The degeneracy between the LOS velocity and microturbulence term usually occurs in the turbulent convection.

We also tried another inversion with only 2 nodes for LOS velocity without the microturbulence term (Appendix \ref{app:A}).
The reduced $\chi^2$ value is 4.47 which is larger than the results with 5 velocity nodes and the deviation is large not only near the line wing but also near the line center.
This implies that the small-scale velocity variation along the LOS direction, which cannot be represented by such a large-scale velocity gradient, exists in the photosphere, although we cannot rule out the effect of the approximate treatment of the collisional broadening in the line wing.
The LOS velocity at approximately $\log\tau=-1$ can be well determined by the Fe I 6301.5 {\AA} and 6302.5 {\AA} lines,
while the LOS velocities at other heights are less constrained.
Multi-line observations may improve the inversion due to the different response functions of the lines.

The Doppler width of spectral line is written as
\begin{align}
\sigma = \sqrt{\frac{2k_BT}{m}+v_t^2}.
\end{align}
A large FWHM of a spectral line can be explained using a high temperature without any microturbulence term.
However, the spectral lines observed in fading granules have large equivalent widths. As the equivalent width has a strong sensitivity to temperature, the low temperature in the middle photosphere can cause a large equivalent width \citep{deltoroiniesta03}. Therefore, explaining both the large FWHM and large equivalent width simultaneously only using the temperature is impossible.
The microturbulence term can simultaneously reproduce both the large FWHM and large equivalent width. The microturbulence term can simultaneously enlarge both the FWHM and equivalent width. The LOS gradient of Doppler velocity can contribute to reproduce the asymmetric profiles of spectral lines.

\section{Comparison with Simulation}
\subsection{Numerical Simulation}
To interpret the inversion result, we check the atmosphere simulated by MURaM code \citep{Vogler05}. In the code, the radiative energy exchange was solved via a non-gray radiative transfer assuming a local thermal equilibrium
(LTE) that reproduced realistic granular scale flows in the photosphere.
The grid size was 10.4 km $\times$ 10.4 $\times$ 14 km,
covering 6 Mm $\times$ 6 Mm $\times$ 1.4 Mm \citep{Riethmuller14}.
A unipolar homogeneous vertical magnetic field of $B_z$ = 30 G
was introduced as an initial condition.
The simulation used a fully-developed hydro-dynamical simulation as the initial condition
and additional calculation for 3 h of solar time was sufficient to reach a statistically stationary state.
The temporal cadence of the data cube is 35 s.
The emergent spectral line profiles of Fe I 6301.5 {\AA} and 6302.5 {\AA} are synthesized with the SIR code \citep{Ruizcobo94}
that calculates a LTE radiative transfer with the line-spread function of the Hinode-SOT/SP \citep{Lites+13}.

\begin{figure*}[tbp]
\begin{center}
\includegraphics[width=15cm]{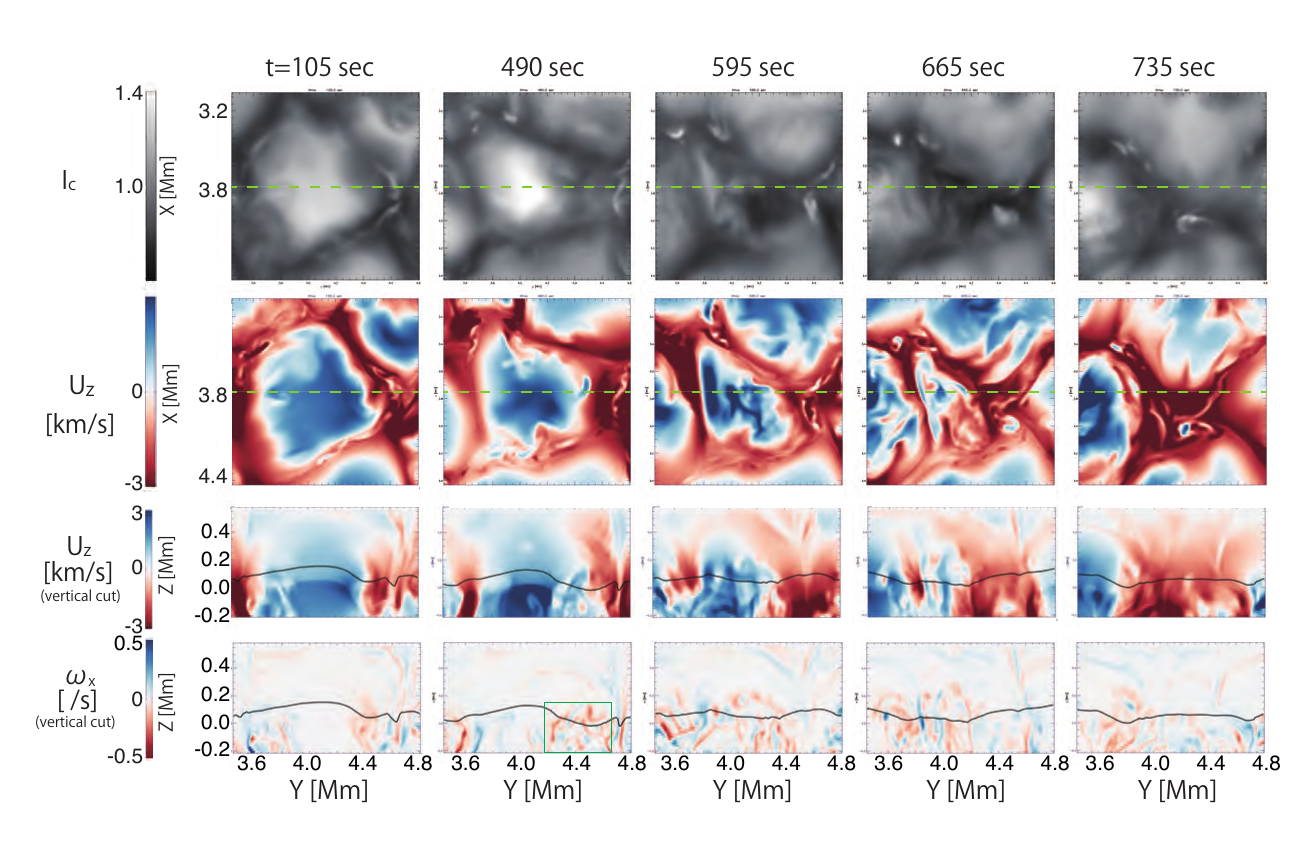}
\end{center}
\caption{Temporal evolution of a fading granule observed in the MURaM simulation data.
  The horizontal distributions of continuum intensity and vertical velocity
 at z=112 km (994 km height from the bottom boundary) are shown in upper rows.
  The vertical cuts at x=4.0 Mm (indicated by the green dashed lines) of vertical velocity
  and the $x-$component of vorticity are shown in bottom rows.
  The black solid lines represent the iso-$\tau$ surfaces of $\log\tau=0$.}
\label{fig:fading_synt1}
\end{figure*}

\begin{figure*}[tbp]
\begin{center}
\includegraphics[width=15cm]{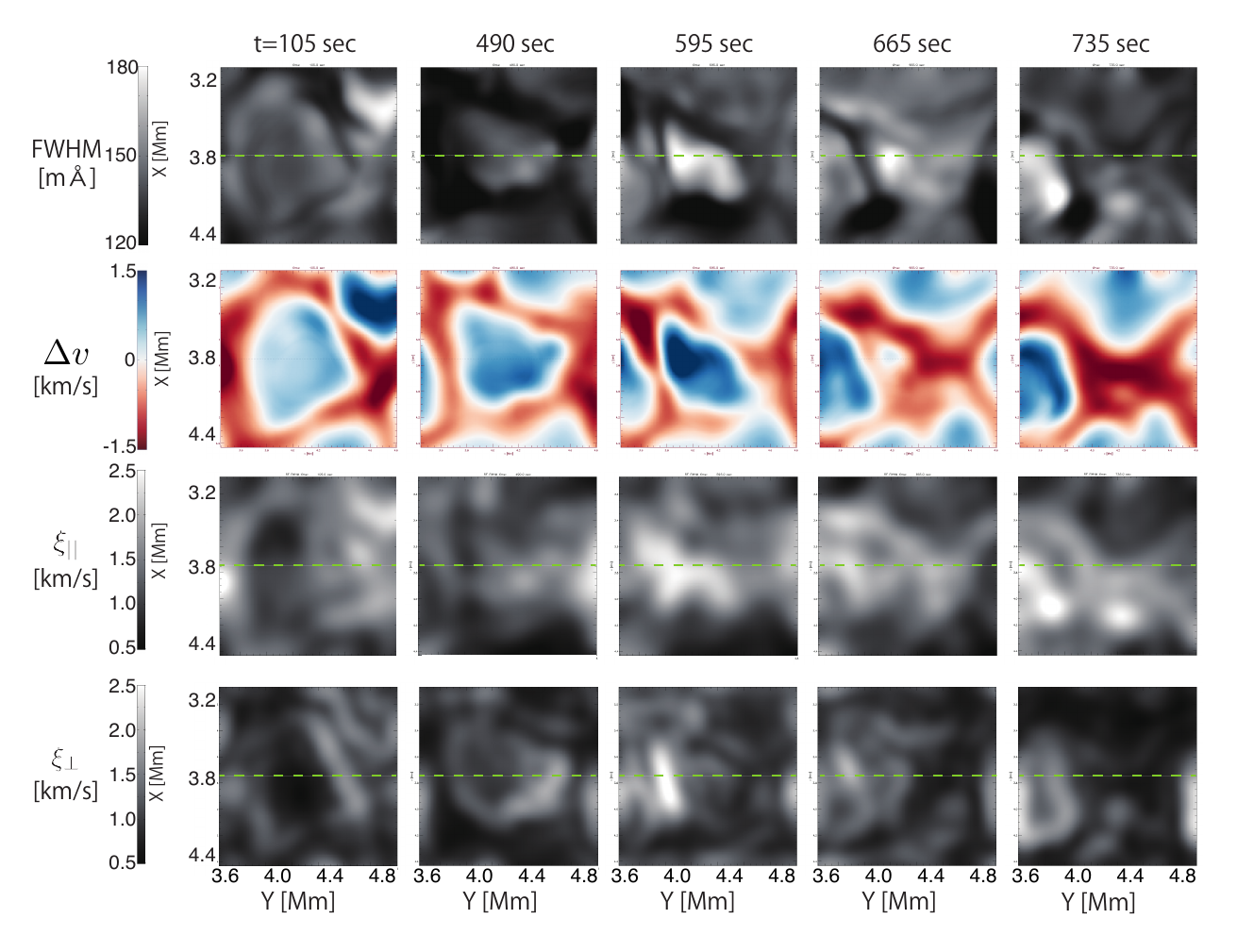}
\end{center}
\caption{
  The horizontal distributions of FWHM and $\Delta v$ with PSF convolution and
  the velocity variances $\xi_{||}$ and $\xi_{\perp}$ calculated for the Fe I 6301.5 {\AA} line (see equations (\ref{eq:turb_para}) and (\ref{eq:turb_perp})) in the same fading granule event in Figure \ref{fig:fading_synt1} are shown.
  }
\label{fig:fading_synt2}
\end{figure*}

\subsection{Evaluation of turbulent velocity}
To quantify the small-scale velocity fields that cannot be resolved by the Hinode-SOT/SP,
we define two velocity variances
using the PSF of the Hinode-SOT/SP
and response function for the FWHM.
The response function for the FWHM associated with the velocity perturbations $R_v^{^{\mathrm{FWHM}}}$ for Fe I 6301.5 {\AA} line is calculated
at each grid using the generalized response function \citep{Ruizcobo94,deltoroiniesta03}.

The generalized response function for the FWHM associated with velocity perturbations is obtained with a linear combination of response function at four wavelength positions around the bisector level of 50\%
as the FWHM is calculated by the linear interpolation of intensities at these four points \citep{RTI20a}.
Hence, we get a simple equation in the same manner that \citet{Gonzalez20} used for the bisector velocities:
\begin{align}
  R_v^{^\mathrm{FWHM}}(\tau) &= \sum_{i} \alpha_i R_v(\lambda_i,\tau),\label{eq:rv_fwhm_def}
\end{align}
where the four coefficients ${\alpha_i}$ for $i\in\{-2, -1, 1, 2\}$ are defined as
\begin{align}
\alpha_i &= \mathrm{sgn}(i)
\frac{(\lambda_i-\lambda_j)(I_j-I_{\mathrm{HM}})}{(I_i-I_j)^2},\label{eq:rv_fwhm_coeff}
\end{align}
where $j=\mathrm{sgn}(i)\left\{|i|+(-1)^{|i|+1}\right\}$ and $\mathrm{sgn}(i)$ is the signum function.
Here, $I_{\mathrm{HM}}$ represents the intensity at the bisector level of 50\% where the FWHM is measured.

Since the coefficients satisfy $\alpha_{1}=\alpha_{-1}$ and $\alpha_{2}=\alpha_{-2}$ and
the response function $R_v$ has an anti-symmetric profile
($R_v(\lambda-\delta\lambda,\tau)=-R_v(\lambda+\delta\lambda,\tau)$),
the response function of FWHM ($R_v^{\mathrm{FWHM}}$) is always zero
for a model atmosphere such as HSRA that does not include any LOS velocity variation along the LOS.
The Doppler velocity term, rather than the microturbulence term, can enlarge the FWHM by the higher-order effect.
The small amplitude perturbation of the LOS velocity profile only causes a small shift or asymmetry
in the spectral line profile and does not lead to line broadening.
In the actual solar atmosphere, the velocity perturbations do not always have small amplitudes so that the response function $R_v^{\mathrm{FWHM}}$ can have a non-zero value.

The response function $R_v^{^{\mathrm{FWHM}}}$ describes
how the LOS velocity over entire atmosphere affects the FWHM.
The vertical average of vertical velocity of each grid is defined as
\begin{align}
\overline{u_z} &= \frac{\int u_z|R_v^{^{\mathrm{FWHM}}}| d\tau}{\int |R_v^{^{\mathrm{FWHM}}}| d\tau}.
\end{align}
and the horizontal average of $\overline{u_z}$ is defined as the convolution with the PSF, $P$:
\begin{align}
\left<\overline{u_z}\right> &= \overline{u_z} * P.
\end{align}
The variances of vertical velocities on the LOS and horizontal direction are then calculated as
\begin{align}
\xi^2_{||} &= 2\left<\overline{(u_z-\overline{u_z})^2}\right> \label{eq:turb_para},\\
\xi^2_{\perp} &= 2\left< (\overline{u_z}-\left<\overline{u_z}\right>)^2\right>. \label{eq:turb_perp}
\end{align}
These definitions are similar to those of \citet{Steffen13} who investigated stellar spectra (observed as a point source),
but considered the PSF of the spatially resolving observation
and response function of 6301.5 {\AA} line.
The LOS variance $\xi^2_{||}$ is defined as the variance of the vertical velocity along the LOS direction,
averaging over the PSF,
and the horizontal variance $\xi^2_{\perp}$ is the variance of the average vertical velocities within the PSF.

Qualitatively, $\xi_{||}$ corresponds to the microturbulence term
while $\xi_{\perp}$ corresponds to the macroturbulence term.
Here $\xi_{||}$ indicates the velocity variance along the LOS direction
on scales either smaller or similar to the height range where the response function exhibits some sensitivity,
averaged over the horizontal scale of PSF.
However, $\xi_{\perp}$ shows the velocity variance along the horizontal direction
on scales smaller than the PSF.

\subsection{Fading granule in the MHD simulation}
Figures \ref{fig:fading_synt1} and \ref{fig:fading_synt2} show an example of a fading granule observed in the simulation.
In this fading granule, the decrease of the continuum intensity and change in the vertical velocity upward to downward 
are observed (Figure \ref{fig:fading_synt1}) and related broad line widths of approximately 180 m{\AA} are reproduced (Figure \ref{fig:fading_synt2}).
The line broadening appears within the spatial scale of 500 km and lasts for about 100 s.
The velocity difference $\Delta v$ also increases when the granule fades out.
These features are consistent with the nature of the observed fading granules described in \citet{RTI20a}.

In the former phase before $t=490$ s, laminar upflows near the center of the granule
are seen in the vertical velocity distribution (Figure \ref{fig:fading_synt1}).
In this phase, vorticity distributions
displayed in the bottom row
are vertically elongated at the boundary between the granule and intergranular lane,
which indicates the existence of velocity shears around the boundary.
Subsequently, the emergent continuum intensity shown in the top row rapidly decreases.
In this fading phase at around 595 s, a significant line broadening occurs as shown in Figure \ref{fig:fading_synt2}.
Small-scale vortices appear inside this fading granule
and the vertical distributions of vertical velocity have small-scale variations as shown in bottom panels in Figure \ref{fig:fading_synt1}.
The enhanced variance of vertical velocity can also be seen
in the spatial distribution of $\xi_{||}$ in Figure \ref{fig:fading_synt2}.
The spatial distributions of large FWHM and large $\xi_{||}$
in the fading granule are in good agreement with each other.
Finally, the fading granule becomes an intergranular lane dominated by strong downflows.
The correlation coefficient between FWHM and $\xi_{||}$
calculated with all the five frames is 0.54,
while that between FWHM and $\xi_{\perp}$ is 0.25.
The FWHM observed with the Hinode resolution
reflects the LOS velocity variance along the LOS direction rather than the horizontal variance.

\begin{figure*}[t]
\begin{center}
\includegraphics[width=14cm]{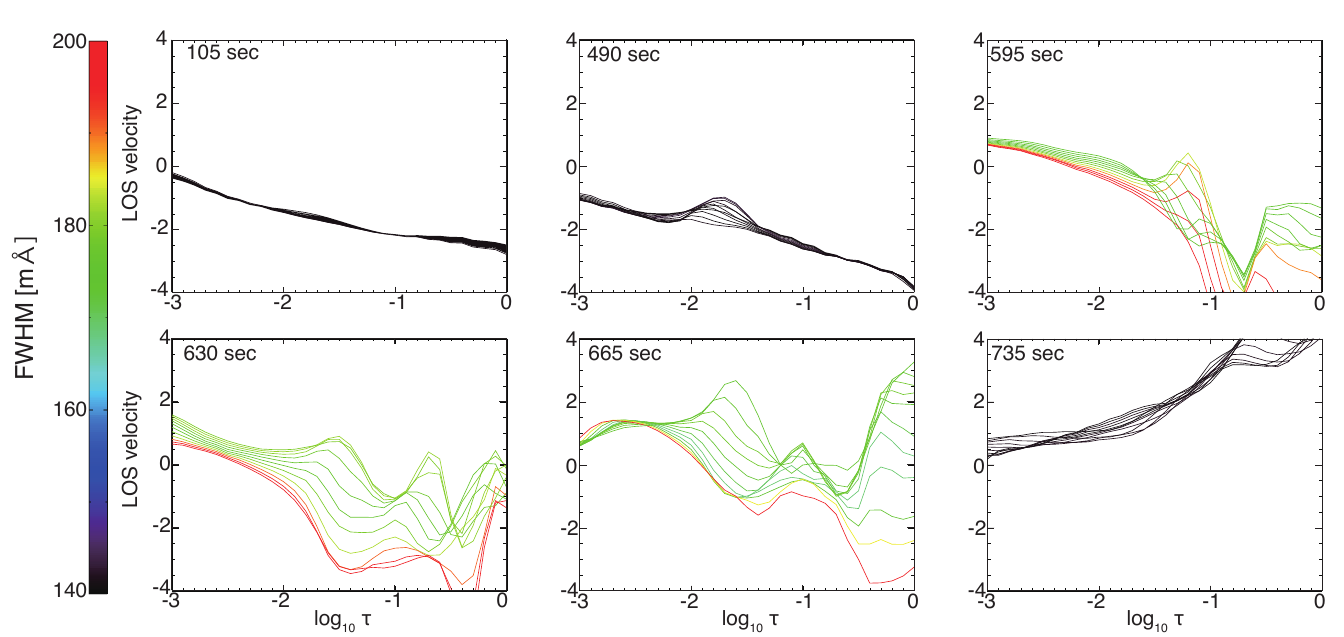}
\end{center}
\caption{The temporal evolutions of the LOS velocities in the fading granule are shown in Figures \ref{fig:fading_synt1} and \ref{fig:fading_synt2}.
  LOS velocities at 11 grids around $Y\in[4.02,4.12]$ Mm along the green line in upper panels in Figure \ref{fig:fading_synt1} and Figure \ref{fig:fading_synt2} are presented.
  The color shows the FWHM at original resolution, i.e. without the convolution of the Hinode PSF.}
\label{fig:muram_vel}
\end{figure*}

Figure \ref{fig:muram_vel} shows the temporal evolutions of the LOS velocity profiles in the fading granule
without the PSF convolution.
LOS velocity profiles at 11 simulation grids that cover the horizontal scale of 100 km are presented.
At $t=105$ s, the LOS velocities are similar to those in the adjacent simulation grids within a scale of 100 km.
The LOS velocities have a velocity gradient that is almost constant along the LOS.
At $t=595$ s, small-scale variations in the velocity gradient exist in the lower photosphere
from $\log\tau=0$ to $-2$.
These small-scale fluctuations correspond to the LOS velocity variation at around $y=4.0$ Mm and $z=0.1$ Mm
shown in the vertical cut in Figure \ref{fig:fading_synt1},
changing the sings of vorticities within 50 km.
These turbulent motions and corresponding line broadenings exist until $t=665$ s as shown in Figure \ref{fig:muram_vel}.
The horizontal scale of these fluctuations is also so small that the LOS velocity profiles differ from the neighboring simulation grids.
At $t=735$ s, the velocity fluctuations remain but with a small amplitude,
consequently, the FWHM is small.

At $t=595$ s,
the localized gradient of vertical velocity
appears in a few grids along the vertical direction in the simulation,
which corresponds to a few 10 km.
This causes a large variance of LOS velocities at $-1.5<\log\tau<-0.5$.
Therefore, this localized gradient can work similarly as the microturbulence term and broaden the line profiles.

\section{Discussion: driving of turbulent flows}\label{sec:discussion_syn}
In the MURaM simulation data, spectral line broadening in a fading granule
is associated with the turbulent motions in the lower photosphere.
A typical granule with a bright continuum intensity
consists of hot and less turbulent upflow plumes
in this numerical simulation.
At the edge of such typical granule, owing to the presence of an intergranular lane,
a horizontal gradient of vertical velocity exists inherently.
\citet{Nesis93} concluded that line broadening can occur in concentrated regions
with a large horizontal gradient of vertical velocity.
\citet{Solanki96} suggested that such a horizontal gradient of vertical velocity
can broaden the spectral line width because of a type of macroturbulence effect.
As suggested by \citet{Solanki96}, the horizontal velocity variance $\xi_{\perp}$, which corresponds to the macroturbulence, is large at the boundary
where the horizontal gradient of vertical velocity exists.
However, the resultant spectral line profile at the boundary
does not have a large FWHM
at approximately $y=3.8$ Mm at t=$595$ s as shown in Figure \ref{fig:fading_synt2},
even though considering the PSF of the Hinode-SOT/SP
that can cover the strong shear flow structure.
\citet{Khomenko10} found both observationally and numerically
that the spectral line profiles with small FWHM are preferentially observed
at the boundary between a granule and an intergranular lane
because the vertical gradient of LOS velocity is small.

Small-scale velocity variations along the LOS can broaden the spectral line widths.
During the fading process of a granule,
small-scale turbulent motions are developed as shown in Figure \ref{fig:fading_synt1}.
Small-scale velocities emphasize the turbulent nature in the fading granule.
These turbulent motions result in small-scale fluctuations of LOS velocity at $-1.5<\log\tau<-0.5$
where the Fe I 6302 lines are formed.
The spatial scale of these fluctuations are smaller than the scales of the response function of the iron lines, so they can broaden the spectral lines.
Both the vertical and horizontal fluctuations of LOS velocity potentially can broaden the spectral line profiles.
The contribution of the vertical structures are demonstrated as shown in Figure \ref{fig:muram_vel}.
However, the contribution of the horizontal structures are not obvious.
As the region with large FWHM (or developed turbulent motions) are localized,
the large FWHM and adjacent small FWHM can be averaged by the PSF.

These turbulent motions can be triggered
by the shear flow or vorticity at the boundary between the granule and  intergranular lane.
Formation of vortex tube at the edges of granules were found observationally \citep{Steiner10}
and numerically \citep{Kitiashvili12}.
Such vortical flows can excite turbulent motions (green rectangle in Figure \ref{fig:fading_synt1})
via the Kelvin-Helmholtz instability.
During the fading process of granules, the turbulent motions grow and intrude into the granules.
The turbulent motions contributing to the line broadening in the fading granule
could be driven by the shear flows of upward and downward velocities
at the boundary between the granule and intergranular lane.
This implies that the excess line broadening
found by \citet{Nesis93} and \citet{Hanslmeier94}
may be related to these turbulent motions.

After the fading process of a granule, converging flows appear
probably due to decrease in the gas pressure.
Although the fading process is not always associated with the magnetic fields
as shown in \citet{RTI20a},
the magnetic flux can be advected and amplified by these converging flows
if it is weak around a fading site.
Such amplification of magnetic fields is found by \citet{Rempel18}
in newly formed downflow lanes
that cause converging flows and subsequent turbulent flows.
As the turbulent motions are developed already
when the magnetic field are advected into the fading site,
the small-scale interaction and energy conversion can occur immediately.


As shown in Figures \ref{fig:inv_fading_time} and \ref{fig:inv_fading_time2},
there are the Doppler velocity variations along the LOS direction, as manifested by an increase in the bisector velocity difference $\Delta v$.
This numerical simulation also suggests the existence of small-scale velocity variations especially in the lower photosphere.
Although the Doppler velocity in the lower photosphere cannot be determined by the Fe I 6302 {\AA} lines,
multi-line observation and simultaneous inversion are a powerful way to probe such a complex velocity distribution.
For example, the neutral iron line at 1.56 {\textmu}m formed in the bottom photosphere \citep{Milic19} can have a strong capability to constrain the velocity profiles.
By observing both the 6302 {\AA} and 1.56 {\textmu}m lines,
we can constrain the large-scale velocity gradient along the LOS
as well as detect the LOS variations of small-scale velocities
via the spectral line widths.

\section{Summary}
We estimated the atmospheric structure associated with the wide spectral line
by performing spectral line inversion with the SIR code.
Two inversion scenarios were considered: with and without microturbulence term.
Microturbulent velocity of approximately 1 km/s can explain the wide spectral line width,
while the line width was also reproduced without microturbulence
by including large gradients of temperature and Doppler velocity.
However, the narrow line profiles in granule and intergranular lanes
can be fitted without the microturbulence term,
which shows the average microturbulence term in the quiet region
is smaller than 1 km/s.
This value is much smaller than the average microturbulent velocity of 3.1 km/s
at the solar surface obtained by \citet{Buehler15}.

We also analyzed the MHD simulation data obtained with the MURaM code.
Turbulent motions are excited in the vicinity of a fading granule,
which broaden the emergent spectral line width.
As the spatial scales of the velocity variation in the lower photosphere
are $\sim 10$ km smaller than $\Delta(\log\tau)\lesssim1$,
which demonstrates that the MHD simulation supports the inversion scenario with the microturbulence term.
We showed that small-scale velocity variations along the LOS direction in the lower photosphere
are a dominant source of the excess broadening,
which indicates that the line broadening
can be a good tracer of turbulent motions in the photosphere.
We cannot completely differentiate between microturbulence and a complex velocity structure only
from the observational and inversion analyses of a single line profile, but MHD simulation favors small-scale
turbulent motion as a cause of line broaening.

To evaluate such small-scale velocity fields by spectral line inversion,
numerous nodes for LOS velocity in the lower photosphere are required. Otherwise, the inclusion of the microturbulence term becomes necessary in the lower layer.
Observing other lines sensitive to the lower photosphere is also important.
Owing to the low continuum opacity,
Fe I 15650 {\AA} lines are good candidates for assessing the LOS velocity in the lower photosphere,
which can be observed with Daniel K. Inouye Solar Telescope (DKIST; \citealt{Rimmele20,Rast21}).
DKIST has a 4-m aperture and the Diffraction-Limited Near-IR SpectroPolarimeter
has a spatial resolution of 0.1 arcsec at 15650 {\AA} corresponding to 70 km on the solar surface.
This can spatially resolve the local region with wide spectral line profile in a fading phase of granule.
The simultaneous spatial and spectral coverage and high temporal cadence observation
enable us to capture the rapid change in the small-scale flow fields and corresponding line profiles.
In addition, observing Fe I 8468 {\AA} and 8514 {\AA} lines
via the SUNRISE Chromospheric Infared spectroPolarimeter \citep{Katsukawa20} onboard
SUNRISE-3 balloon-borne telescope \citep{Feller20},
enables us to determine the LOS velocity in the upper photosphere.

\section*{}
Hinode is a Japanese mission developed and launched by ISAS/JAXA,
in collaboration with NAOJ as the domestic partner, and NASA and STFC as the international partners.
The support for post-launch operation is provided by JAXA and NAOJ (Japan),
STFC (UK), NASA, ESA, and NSC (Norway).
R.T.I. is supported by JSPS Research Fellowships for Young Scientists.
This study is supported by the JSPS KAKENHI Grant Numbers JP18H05234 and JP23K25916 (PI: Y. Katsukawa), JP19J20294 (PI: R.T. Ishikawa), 23KJ0299 (PI: R.T. Ishikawa), and 20KK0072 (PI: S. Toriumi).
The authors acknowledge financial support by the NINS program for cross-disciplinary study
(Grant Numbers 01321802 and 01311904) on Turbulence, Transport, and Heating Dynamics
in Laboratory and Solar/Astrophysical Plasmas: "SoLaBo-X".

\appendix
\section{Inversion with 2 velocity nodes}\label{app:A}
\begin{figure*}[t]
\begin{center}
\includegraphics[width=14cm]{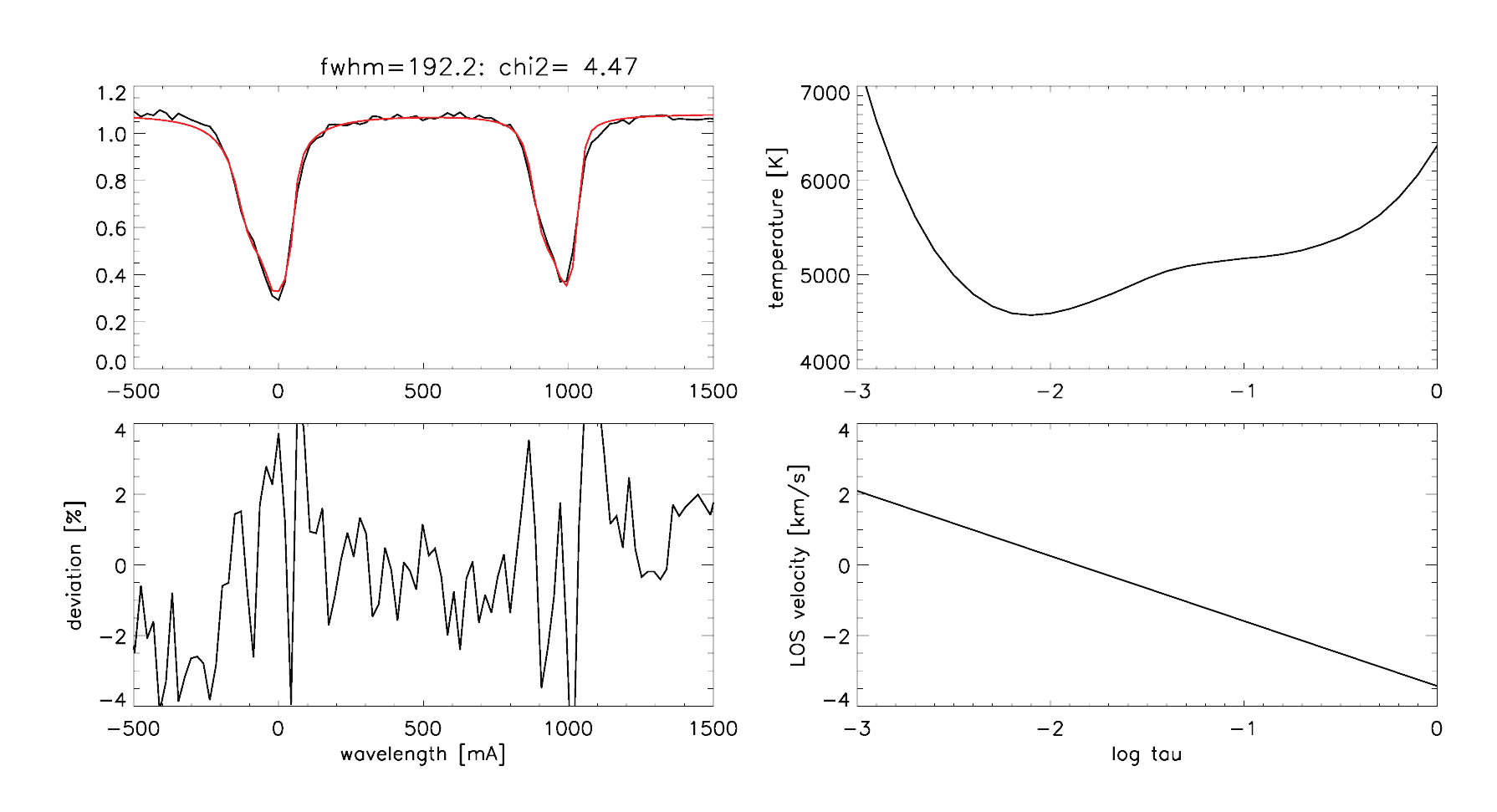}
\end{center}
\caption{The inversion result with 2 velocity nodes and 5 temperature nodes without microturbulence term. The spectral line profile analyzed here is the same as that shown in Figure \ref{fig:inv_fading_time}.}
\label{fig:app_2velnode}
\end{figure*}

We conducted a spectral line inversion with 2 velocity nodes, instead of 5 nodes used in Section \ref{sec:method_inversion}, without the microturbulence term. The number of nodes for temperature is 5. 
Figure \ref{fig:app_2velnode} shows the result.
The spectral line profile analyzed here is the same as that shown in Figure \ref{fig:inv_fading_time}.
The inversion result shows a larger reduced $\chi^2$ compared with the results with 5 velocity nodes shown in Figure \ref{fig:inv_fading_time}.
The deviation is large especially near the line core.
This result implies the existence of small-scale velocity variations along the LOS direction.

\bibliography{ref2}{}

\begin{thebibliography}{}
\expandafter\ifx\csname natexlab\endcsname\relax\def\natexlab#1{#1}\fi
\providecommand{\url}[1]{\href{#1}{#1}}
\providecommand{\dodoi}[1]{doi:~\href{http://doi.org/#1}{\nolinkurl{#1}}}
\providecommand{\doeprint}[1]{\href{http://ascl.net/#1}{\nolinkurl{http://ascl.net/#1}}}
\providecommand{\doarXiv}[1]{\href{https://arxiv.org/abs/#1}{\nolinkurl{https://arxiv.org/abs/#1}}}

\bibitem[{{Abramenko} {et~al.}(2012){Abramenko}, {Yurchyshyn}, {Goode}, {Kitiashvili}, \& {Kosovichev}}]{Abramenko12}
{Abramenko}, V.~I., {Yurchyshyn}, V.~B., {Goode}, P.~R., {Kitiashvili}, I.~N., \& {Kosovichev}, A.~G. 2012, \apjl, 756, L27.
\newblock \doarXiv{1208.4337}

\bibitem[{{Bellot Rubio}(2009)}]{BellotRubio09}
{Bellot Rubio}, L.~R. 2009, \apj, 700, 284, \dodoi{10.1088/0004-637X/700/1/284}

\bibitem[{{Bellot Rubio} \& {Beck}(2005)}]{BellotRubio05}
{Bellot Rubio}, L.~R., \& {Beck}, C. 2005, \apjl, 626, L125, \dodoi{10.1086/431648}

\bibitem[{{Buehler} {et~al.}(2015){Buehler}, {Lagg}, {Solanki}, \& {van Noort}}]{Buehler15}
{Buehler}, D., {Lagg}, A., {Solanki}, S.~K., \& {van Noort}, M. 2015, \aap, 576, A27, \dodoi{10.1051/0004-6361/201424970}

\bibitem[{{del Toro Iniesta}(2003)}]{deltoroiniesta03}
{del Toro Iniesta}, J.~C. 2003, {Introduction to Spectropolarimetry}

\bibitem[{{Feller} {et~al.}(2020){Feller}, {Gandorfer}, {Iglesias}, {Lagg}, {Riethm{\"u}ller}, {Solanki}, {Katsukawa}, \& {Kubo}}]{Feller20}
{Feller}, A., {Gandorfer}, A., {Iglesias}, F.~A., {et~al.} 2020, in Society of Photo-Optical Instrumentation Engineers (SPIE) Conference Series, Vol. 11447, Society of Photo-Optical Instrumentation Engineers (SPIE) Conference Series, 11447AK, \dodoi{10.1117/12.2562666}

\bibitem[{{Gadun} {et~al.}(1997){Gadun}, {Hanslmeier}, \& {Pikalov}}]{Gadun97}
{Gadun}, A.~S., {Hanslmeier}, A., \& {Pikalov}, K.~N. 1997, \aap, 320, 1001

\bibitem[{{Gingerich} {et~al.}(1971){Gingerich}, {Noyes}, {Kalkofen}, \& {Cuny}}]{Gingerich71}
{Gingerich}, O., {Noyes}, R.~W., {Kalkofen}, W., \& {Cuny}, Y. 1971, \solphys, 18, 347, \dodoi{10.1007/BF00149057}

\bibitem[{{Gonz{\'a}lez Manrique} {et~al.}(2020){Gonz{\'a}lez Manrique}, {Quintero Noda}, {Kuckein}, {Ruiz Cobo}, \& {Carlsson}}]{Gonzalez20}
{Gonz{\'a}lez Manrique}, S.~J., {Quintero Noda}, C., {Kuckein}, C., {Ruiz Cobo}, B., \& {Carlsson}, M. 2020, \aap, 634, A19, \dodoi{10.1051/0004-6361/201937274}

\bibitem[{{Guglielmino} {et~al.}(2020){Guglielmino}, {Mart{\'\i}nez Pillet}, {Ruiz Cobo}, {Bellot Rubio}, {del Toro Iniesta}, {Solanki}, {Riethm{\"u}ller}, \& {Zuccarello}}]{Guglielmino20}
{Guglielmino}, S.~L., {Mart{\'\i}nez Pillet}, V., {Ruiz Cobo}, B., {et~al.} 2020, \apj, 896, 62, \dodoi{10.3847/1538-4357/ab917b}

\bibitem[{{Hanslmeier} {et~al.}(1994){Hanslmeier}, {Nesis}, \& {Mattig}}]{Hanslmeier94}
{Hanslmeier}, A., {Nesis}, A., \& {Mattig}, W. 1994, \aap, 288, 960

\bibitem[{{Hirzberger} {et~al.}(1999){Hirzberger}, {Bonet}, {V{\'a}zquez}, \& {Hanslmeier}}]{Hirzberger99a}
{Hirzberger}, J., {Bonet}, J.~A., {V{\'a}zquez}, M., \& {Hanslmeier}, A. 1999, \apj, 515, 441

\bibitem[{{Holweger} \& {Kneer}(1989)}]{Holweger89}
{Holweger}, H., \& {Kneer}, F. 1989, in NATO Advanced Study Institute (ASI) Series C, Vol. 263, Solar and Stellar Granulation, ed. R.~J. {Rutten} \& G.~{Severino}, 173

\bibitem[{{Ishikawa} {et~al.}(2020){Ishikawa}, {Katsukawa}, {Oba}, {Nakata}, {Nagaoka}, \& {Kobayashi}}]{RTI20a}
{Ishikawa}, R.~T., {Katsukawa}, Y., {Oba}, T., {et~al.} 2020, \apj, 890, 138, \dodoi{10.3847/1538-4357/ab6bce}

\bibitem[{{Katsukawa} \& {Orozco Su{\'a}rez}(2012)}]{Katsukawa12}
{Katsukawa}, Y., \& {Orozco Su{\'a}rez}, D. 2012, \apj, 758, 139.
\newblock \doarXiv{1209.0548}

\bibitem[{{Katsukawa} {et~al.}(2020){Katsukawa}, {del Toro Iniesta}, {Solanki}, {Kubo}, {Hara}, {Shimizu}, {Oba}, {Kawabata}, {Tsuzuki}, {Uraguchi}, {Nodomi}, {Shinoda}, {Tamura}, {Suematsu}, {Ishikawa}, {Kano}, {Matsumoto}, {Ichimoto}, {Nagata}, {Quintero Noda}, {Anan}, {Orozco Su{\'a}rez}, {Balaguer Jim{\'e}nez}, {L{\'o}pez Jim{\'e}nez}, {Cobos Carrascosa}, {Feller}, {Riethmueller}, {Gandorfer}, \& {Lagg}}]{Katsukawa20}
{Katsukawa}, Y., {del Toro Iniesta}, J.~C., {Solanki}, S.~K., {et~al.} 2020, in Society of Photo-Optical Instrumentation Engineers (SPIE) Conference Series, Vol. 11447, Society of Photo-Optical Instrumentation Engineers (SPIE) Conference Series, 114470Y, \dodoi{10.1117/12.2561223}

\bibitem[{{Khomenko} {et~al.}(2010){Khomenko}, {Mart{\'{\i}}nez Pillet}, {Solanki}, {del Toro Iniesta}, {Gandorfer}, {Bonet}, {Domingo}, {Schmidt}, {Barthol}, \& {Kn{\"o}lker}}]{Khomenko10}
{Khomenko}, E., {Mart{\'{\i}}nez Pillet}, V., {Solanki}, S.~K., {et~al.} 2010, \apjl, 723, L159.
\newblock \doarXiv{1008.0517}

\bibitem[{{Kitiashvili} {et~al.}(2012){Kitiashvili}, {Kosovichev}, {Mansour}, {Lele}, \& {Wray}}]{Kitiashvili12}
{Kitiashvili}, I.~N., {Kosovichev}, A.~G., {Mansour}, N.~N., {Lele}, S.~K., \& {Wray}, A.~A. 2012, \physscr, 86, 018403, \dodoi{10.1088/0031-8949/86/01/018403}

\bibitem[{{Kosugi} {et~al.}(2007){Kosugi}, {Matsuzaki}, {Sakao}, {Shimizu}, {Sone}, {Tachikawa}, {Hashimoto}, {Minesugi}, {Ohnishi}, {Yamada}, {Tsuneta}, {Hara}, {Ichimoto}, {Suematsu}, {Shimojo}, {Watanabe}, {Shimada}, {Davis}, {Hill}, {Owens}, {Title}, {Culhane}, {Harra}, {Doschek}, \& {Golub}}]{Kosugi07}
{Kosugi}, T., {Matsuzaki}, K., {Sakao}, T., {et~al.} 2007, \solphys, 243, 3

\bibitem[{{Landi Degl'Innocenti} \& {Landi Degl'Innocenti}(1977)}]{Landi77}
{Landi Degl'Innocenti}, E., \& {Landi Degl'Innocenti}, M. 1977, \aap, 56, 111

\bibitem[{{Landi Degl'Innocenti} \& {Landi Degl'Innocenti}(1985)}]{Landi85}
---. 1985, \solphys, 97, 239, \dodoi{10.1007/BF00165988}

\bibitem[{{Lemmerer} {et~al.}(2017){Lemmerer}, {Hanslmeier}, {Muthsam}, \& {Piantschitsch}}]{Lemmerer17}
{Lemmerer}, B., {Hanslmeier}, A., {Muthsam}, H., \& {Piantschitsch}, I. 2017, \aap, 598, A126.
\newblock \doarXiv{1611.06786}

\bibitem[{{Lites} \& {Ichimoto}(2013)}]{Lites_Ichimoto13}
{Lites}, B.~W., \& {Ichimoto}, K. 2013, \solphys, 283, 601

\bibitem[{{Lites} {et~al.}(2013){Lites}, {Akin}, {Card}, {Cruz}, {Duncan}, {Edwards}, {Elmore}, {Hoffmann}, {Katsukawa}, {Katz}, {Kubo}, {Ichimoto}, {Shimizu}, {Shine}, {Streander}, {Suematsu}, {Tarbell}, {Title}, \& {Tsuneta}}]{Lites+13}
{Lites}, B.~W., {Akin}, D.~L., {Card}, G., {et~al.} 2013, \solphys, 283, 579

\bibitem[{{L{\'o}pez Ariste} {et~al.}(2007){L{\'o}pez Ariste}, {Mart{\'\i}nez Gonz{\'a}lez}, \& {Ram{\'\i}rez V{\'e}lez}}]{LopezAriste07}
{L{\'o}pez Ariste}, A., {Mart{\'\i}nez Gonz{\'a}lez}, M.~J., \& {Ram{\'\i}rez V{\'e}lez}, J.~C. 2007, \aap, 464, 351, \dodoi{10.1051/0004-6361:20065593}

\bibitem[{{Mili{\'c}} {et~al.}(2019){Mili{\'c}}, {Smitha}, \& {Lagg}}]{Milic19}
{Mili{\'c}}, I., {Smitha}, H.~N., \& {Lagg}, A. 2019, \aap, 630, A133, \dodoi{10.1051/0004-6361/201935126}

\bibitem[{{Nesis} {et~al.}(1992){Nesis}, {Bogdan}, {Cattaneo}, {Hanslmeier}, {Knoelker}, \& {Malagoli}}]{Nesis92}
{Nesis}, A., {Bogdan}, T.~J., {Cattaneo}, F., {et~al.} 1992, \apjl, 399, L99, \dodoi{10.1086/186616}

\bibitem[{{Nesis} {et~al.}(1993){Nesis}, {Hanslmeier}, {Hammer}, {Komm}, {Mattig}, \& {Staiger}}]{Nesis93}
{Nesis}, A., {Hanslmeier}, A., {Hammer}, R., {et~al.} 1993, \aap, 279, 599

\bibitem[{{Oba} {et~al.}(2017){Oba}, {Riethm{\"u}ller}, {Solanki}, {Iida}, {Quintero Noda}, \& {Shimizu}}]{Oba17b}
{Oba}, T., {Riethm{\"u}ller}, T.~L., {Solanki}, S.~K., {et~al.} 2017, \apj, 849, 7.
\newblock \doarXiv{1709.06933}

\bibitem[{{Quintero Noda} {et~al.}(2014){Quintero Noda}, {Borrero}, {Orozco Su{\'a}rez}, \& {Ruiz Cobo}}]{QuinteroNoda14b}
{Quintero Noda}, C., {Borrero}, J.~M., {Orozco Su{\'a}rez}, D., \& {Ruiz Cobo}, B. 2014, \aap, 569, A73, \dodoi{10.1051/0004-6361/201424131}

\bibitem[{{Rast} {et~al.}(2021){Rast}, {Bello Gonz{\'a}lez}, {Bellot Rubio}, {Cao}, {Cauzzi}, {Deluca}, {de Pontieu}, {Fletcher}, {Gibson}, {Judge}, {Katsukawa}, {Kazachenko}, {Khomenko}, {Landi}, {Mart{\'\i}nez Pillet}, {Petrie}, {Qiu}, {Rachmeler}, {Rempel}, {Schmidt}, {Scullion}, {Sun}, {Welsch}, {Andretta}, {Antolin}, {Ayres}, {Balasubramaniam}, {Ballai}, {Berger}, {Bradshaw}, {Campbell}, {Carlsson}, {Casini}, {Centeno}, {Cranmer}, {Criscuoli}, {Deforest}, {Deng}, {Erd{\'e}lyi}, {Fedun}, {Fischer}, {Gonz{\'a}lez Manrique}, {Hahn}, {Harra}, {Henriques}, {Hurlburt}, {Jaeggli}, {Jafarzadeh}, {Jain}, {Jefferies}, {Keys}, {Kowalski}, {Kuckein}, {Kuhn}, {Kuridze}, {Liu}, {Liu}, {Longcope}, {Mathioudakis}, {McAteer}, {McIntosh}, {McKenzie}, {Miralles}, {Morton}, {Muglach}, {Nelson}, {Panesar}, {Parenti}, {Parnell}, {Poduval}, {Reardon}, {Reep}, {Schad}, {Schmit}, {Sharma}, {Socas-Navarro}, {Srivastava}, {Sterling}, {Suematsu}, {Tarr}, {Tiwari}, {Tritschler}, {Verth}, {Vourlidas}, {Wang}, {Wang}, {NSO and DKIST
  Project}, {DKIST Instrument Scientists}, {DKIST Science Working Group}, \& {DKIST Critical Science Plan Community}}]{Rast21}
{Rast}, M.~P., {Bello Gonz{\'a}lez}, N., {Bellot Rubio}, L., {et~al.} 2021, \solphys, 296, 70, \dodoi{10.1007/s11207-021-01789-2}

\bibitem[{{Rempel}(2014)}]{Rempel14}
{Rempel}, M. 2014, \apj, 789, 132.
\newblock \doarXiv{1405.6814}

\bibitem[{{Rempel}(2018)}]{Rempel18}
---. 2018, \apj, 859, 161, \dodoi{10.3847/1538-4357/aabba0}

\bibitem[{{Riethm{\"u}ller} {et~al.}(2014){Riethm{\"u}ller}, {Solanki}, {Berdyugina}, {Sch{\"u}ssler}, {Mart{\'\i}nez Pillet}, {Feller}, {Gandorfer}, \& {Hirzberger}}]{Riethmuller14}
{Riethm{\"u}ller}, T.~L., {Solanki}, S.~K., {Berdyugina}, S.~V., {et~al.} 2014, \aap, 568, A13, \dodoi{10.1051/0004-6361/201423892}

\bibitem[{{Rieutord} {et~al.}(2010){Rieutord}, {Roudier}, {Rincon}, {Malherbe}, {Meunier}, {Berger}, \& {Frank}}]{Rieutord10}
{Rieutord}, M., {Roudier}, T., {Rincon}, F., {et~al.} 2010, \aap, 512, A4, \dodoi{10.1051/0004-6361/200913303}

\bibitem[{{Rimmele} {et~al.}(2020){Rimmele}, {Warner}, {Keil}, {Goode}, {Kn{\"o}lker}, {Kuhn}, {Rosner}, {McMullin}, {Casini}, {Lin}, {W{\"o}ger}, {von der L{\"u}he}, {Tritschler}, {Davey}, {de Wijn}, {Elmore}, {Fehlmann}, {Harrington}, {Jaeggli}, {Rast}, {Schad}, {Schmidt}, {Mathioudakis}, {Mickey}, {Anan}, {Beck}, {Marshall}, {Jeffers}, {Oschmann}, {Beard}, {Berst}, {Cowan}, {Craig}, {Cross}, {Cummings}, {Donnelly}, {de Vanssay}, {Eigenbrot}, {Ferayorni}, {Foster}, {Galapon}, {Gedrites}, {Gonzales}, {Goodrich}, {Gregory}, {Guzman}, {Guzzo}, {Hegwer}, {Hubbard}, {Hubbard}, {Johansson}, {Johnson}, {Liang}, {Liang}, {McQuillen}, {Mayer}, {Newman}, {Onodera}, {Phelps}, {Puentes}, {Richards}, {Rimmele}, {Sekulic}, {Shimko}, {Simison}, {Smith}, {Starman}, {Sueoka}, {Summers}, {Szabo}, {Szabo}, {Wampler}, {Williams}, \& {White}}]{Rimmele20}
{Rimmele}, T.~R., {Warner}, M., {Keil}, S.~L., {et~al.} 2020, \solphys, 295, 172, \dodoi{10.1007/s11207-020-01736-7}

\bibitem[{{Ruiz Cobo} \& {del Toro Iniesta}(1992)}]{Ruizcobo92}
{Ruiz Cobo}, B., \& {del Toro Iniesta}, J.~C. 1992, \apj, 398, 375

\bibitem[{{Ruiz Cobo} \& {del Toro Iniesta}(1994)}]{Ruizcobo94}
---. 1994, \aap, 283, 129

\bibitem[{{Sanchez Almeida}(1992)}]{SanchezAlmeida92}
{Sanchez Almeida}, J. 1992, \solphys, 137, 1, \dodoi{10.1007/BF00146572}

\bibitem[{{Socas-Navarro}(2011)}]{Socasnavarro11}
{Socas-Navarro}, H. 2011, \aap, 529, A37, \dodoi{10.1051/0004-6361/201015805}

\bibitem[{{Solanki} {et~al.}(1996){Solanki}, {Rueedi}, {Bianda}, \& {Steffen}}]{Solanki96}
{Solanki}, S.~K., {Rueedi}, I., {Bianda}, M., \& {Steffen}, M. 1996, \aap, 308, 623

\bibitem[{{Steffen} {et~al.}(2013){Steffen}, {Caffau}, \& {Ludwig}}]{Steffen13}
{Steffen}, M., {Caffau}, E., \& {Ludwig}, H.~G. 2013, Memorie della Societa Astronomica Italiana Supplementi, 24, 37.
\newblock \doarXiv{1306.4307}

\bibitem[{{Steiner} {et~al.}(2010){Steiner}, {Franz}, {Bello Gonz{\'a}lez}, {Nutto}, {Rezaei}, {Mart{\'\i}nez Pillet}, {Bonet Navarro}, {del Toro Iniesta}, {Domingo}, {Solanki}, {Kn{\"o}lker}, {Schmidt}, {Barthol}, \& {Gandorfer}}]{Steiner10}
{Steiner}, O., {Franz}, M., {Bello Gonz{\'a}lez}, N., {et~al.} 2010, \apjl, 723, L180, \dodoi{10.1088/2041-8205/723/2/L180}

\bibitem[{{Tsuneta} {et~al.}(2008){Tsuneta}, {Ichimoto}, {Katsukawa}, {Nagata}, {Otsubo}, {Shimizu}, {Suematsu}, {Nakagiri}, {Noguchi}, {Tarbell}, {Title}, {Shine}, {Rosenberg}, {Hoffmann}, {Jurcevich}, {Kushner}, {Levay}, {Lites}, {Elmore}, {Matsushita}, {Kawaguchi}, {Saito}, {Mikami}, {Hill}, \& {Owens}}]{Tsuneta08}
{Tsuneta}, S., {Ichimoto}, K., {Katsukawa}, Y., {et~al.} 2008, \solphys, 249, 167.
\newblock \doarXiv{0711.1715}

\bibitem[{{Vitas} {et~al.}(2011){Vitas}, {Fischer}, {V{\"o}gler}, \& {Keller}}]{Vitas11}
{Vitas}, N., {Fischer}, C.~E., {V{\"o}gler}, A., \& {Keller}, C.~U. 2011, \aap, 532, A110, \dodoi{10.1051/0004-6361/201015773}

\bibitem[{{V{\"o}gler} {et~al.}(2005){V{\"o}gler}, {Shelyag}, {Sch{\"u}ssler}, {Cattaneo}, {Emonet}, \& {Linde}}]{Vogler05}
{V{\"o}gler}, A., {Shelyag}, S., {Sch{\"u}ssler}, M., {et~al.} 2005, \aap, 429, 335

\end{thebibliography}
\bibliographystyle{aasjournal}



\end{document}